\documentclass[12pt,preprint]{aastex}
\usepackage{graphicx}
\begin{document}
\def\la{\mathrel{\hbox{\rlap{\hbox{\lower4pt\hbox{$\sim$}}}\hbox{$<$}}}}
\def\ga{\mathrel{\hbox{\rlap{\hbox{\lower4pt\hbox{$\sim$}}}\hbox{$>$}}}}
\def\lam{$\lambda$}
\def\kms{km~s$^{-1}$}
\def\vphot{$v_{\rm phot}$}
\def\ang{~\AA}
\def\syn{SYNOW}
\def\dm15{{$\Delta$}$m_{15}$}
\def\rsi{$R$(Si~II)}
\def\v10{$V_{10}$(Si~II)}
\def\wsi{$W_lambda$(Si~II)}
\def\vdot{$\.v$(Si~II)}
\def\W575{$W(5750)$}
\def\W610{$W(6100)$}
\def\6100{the 6100~\AA\ absorption}
\def\tex{$T_{\rm exc}$}
\def\ve{$v_{\rm e}$}

\title {Comparative Direct Analysis of Type~Ia Supernova Spectra. \\
V. Insights from a Larger Sample and Quantitative Subclassification}

\author {David Branch, Leeann Chau Dang, \& E.~Baron}

\affil {Homer L. Dodge Department of Physics and Astronomy, University
of Oklahoma, Norman,~OK; branch@nhn.ou.edu}

\begin{abstract}

A comparative study of optical spectra of Type~Ia supernovae (SNe~Ia)
is extended, in the light of new data.  The discussion is framed in
terms of the four groups defined in previous papers of this series:
core normal (CN); broad line (BL); cool (CL); and shallow silicon
(SS).  Emerging features of the SN~Ia spectroscopic diversity include
evidence (1) that extreme CL SN~1991bg--likes are not a physically
distinct subgroup and (2) for the existence of a substantial number of
SN~1999aa--like SSs that are very similar to each other and
distinguishable from CN even as late as three weeks after maximum
light.  SN~1999aa--likes may be relatively numerous, yet not a
physically distinct subgroup.  The efficacy of quantitative
spectroscopic subclassification of SNe~Ia based on the equivalent
widths of absorption features near 5750\ang\ and 6100\ang\ near
maximum light is discussed.  The absolute magnitude dispersion of a
small sample of CNs is no larger than the characteristic absolute
magnitude uncertainty.

\end{abstract}

\keywords{supernovae: general}

\section{INTRODUCTION}

This is the fifth in a series of papers on a comparative direct
analysis of the optical spectra of Type~Ia supernovae (SNe~Ia).
Papers~I--IV (Branch et~al. 2005, 2006, 2007, 2008) were concerned
with a time series of spectra of the spectroscopically normal
SN~1994D, and with samples of SN~Ia spectra near maximum light, prior
to maximum light, and after maximum light, respectively.  In those
papers, the parameterized supernova synthetic spectrum code SYNOW was
used to study line identifications.  Since those papers were written,
the sample of spectra available to us has increased substantially,
thanks especially to a large number of spectra published by Matheson
et~al. (2008).  The purposes of this paper are to examine
similarities, differences, and relationships among the spectra of the
expanded sample, and to consider the efficacy of SN~Ia quantitative
spectroscopic subclassification based on the equivalent widths of
absorption features near 5750\ang\ and 6100\ang.

\section{THE CURRENT SAMPLE}

As in the previous papers, we confine our attention to optical
spectra, from the Ca~II H and K feature in the blue ($\sim3700$\ang)
to the Ca~II infrared triplet (Ca~II IR3) in the red ($\sim9000$\ang).
All spectra have been corrected for the redshifts of the host
galaxies, and flattened according to the local normalization technique
of Jeffery et~al. (2007).  Mild smoothing has been applied to some of
the spectra.  The current sample of 186 spectra of 65 SNe~Ia (Table~1)
is restricted to the following ranges of epochs with respect to the
time of maximum light in the B band: an early sample (day~$-11$ or
earlier); a 1 week premaximum sample (day~$-8$ to day~$-6$); a maximum
light sample (day~$-3$ to day~+3); a 1 week postmaximum sample (day~+6
to day~+8); a 3 weeks postmaximum sample (day~+19 to day~+23); and a 3
months postmaximum sample (day~+80 to day~+100).  The only difference
from previous papers of this series is that now, with more spectra
available, the 1 week premaximum sample is from day~$-8$ to day~$-6$,
rather than from day~$-9$ to day~$-5$ as in Paper~III.

\section{FOUR GROUPS}

In Paper~II we divided the maximum light spectra of SNe~Ia into four
groups: core normal (CN), broad line (BL), cool (CL), and shallow
silicon (SS).  The group assignments were made on the basis of
measurements of $W(5750)$ and $W(6100)$, the (pseudo) equivalent
widths of absorption features near 5750\ang\ and 6100\ang, as well as
on the appearance (depth, width, and shape) of the
6100\ang\ absorption.  The 6100\ang\ absorption is produced by the
Si~II \lam6355 transition and the 5750\ang\ absorption usually is
attributed to Si~II \lam5972.  A comparison of the 5750\ang\ and
6100\ang\ absorptions in maximum light spectra of the four groups is
shown in Figure~1.  The extreme BL SN~2006X has a very broad
6100\ang\ absorption but only a weak 5750\ang\ absorption; the
6100\ang\ absorption of the extreme CL SN~1991bg is comparable to that
of the CN SN~1998bu (except less blueshifted) but the
5750\ang\ absorption is strong; and the extreme SS SN~1991T has only a
shallow 6100\ang\ absorption and hardly any 5750\ang\ absorption.
(SN~1991T would be less different from the others if a spectrum closer
to maximum light was available.)  In Papers II--IV the presentation
and discussion were in terms of the four groups, although it was
suggested that for the most part the spectra appeared to have a
continuous distribution of properties, rather than breaking up into
discrete subgroups --- extreme SS SN~2002cx--likes being an apparent
exception, and extreme CL SN~1991bg--likes being a possible exception.

In Paper~II $W(5750)$ was plotted against $W(6100)$ for the 24 SNe of
the maximum light sample.  For the present paper all equivalent widths
(Table~1) have been measured in a manner similar to that described by
Hachinger et~al. (2006).  The flattening of the spectra has no
significant effect on our measurements.  Since each set of equivalent
widths reported by us are for a specific spectrum obtained within 3
days of maximum light, while those reported by Hachinger et~al. are
based on multiple spectra obtained within 5 days of maximum light and
extrapolated or interpolated to maximum light, exact agreement is not
expected.  For the 22 SNe in both of our samples, the mean absolute
differences in $W(5750)$ and $W(6100)$ are 2.7\ang\ and 4.7\ang,
respectively.  

Figure~2 shows our current equivalent width plot, for 59 SNe.  In
general Figure~2 resembles its counterpart in Paper~II, but now with
more than double the data.  The plot is most densely populated in the
CN region, with members of the other groups more widely dispersed.
For the present sample the domains of the four groups do not overlap,
except for the maverick CL SN~1989B, which will be discussed in
\S~6. Features that were not present in the plot of Paper~II include
(1) the presence of other CLs in the vicinity of SN~1986G, which had
been rather isolated before, and (2) the clump of four SSs near
$W(6100)\simeq$50\ang, $W(5750)\simeq$8\ang.  These two developments
will be discussed in \S~6 and \S~7.

Figure~3 is like Figure~2 but for the 28 SNe for which Hachinger
et~al. report $W(5750)$ and $W(6100)$ values [$EW (F4)$ and $EW (F5)$
  in their notation], and with our group symbols.  The distribution of
SNe in Figure~3 is much like that of Figure~2 although with less data.

Figure~4 is like Figure~2, but also showing $W(5750)$ and $W(6100)$
values measured from synthetic spectra for the Si~II ion, generated
with the SYNOW code. (SYNOW is discussed in previous papers of this
series.  The parameter $v_{phot}$ is the velocity at the photosphere;
$v_e$ is the e--folding velocity of the line optical depths; $v_{max}$
is an optional maximum velocity of the line--forming region, and
$T_{exc}$ is the excitation temperature used to determine the relative
optical depths of lines of an ion.)  Several example Si~II synthetic
spectra are shown in Figure~5.  The results in Figure~4 are generally
consistent with how we fitted maximum light spectra in Paper~II.  We
used $v_e=1000$ \kms\ for CNs and larger values for BLs.  For CLs we
imposed maximum velocities on Si~II, in order to saturate the
$W(6100)$ values and obtain high values of $W(5750)$ by means of high
line optical depth\footnote{Including the bluer component of the
  5750\ang\ blend in CLs (see Figure~1) provides a good empirical
  separation of CLs and BLs in Figure~2, but it should be noted that
  the bluer component is not accounted for by Si~II \lam5972 in SYNOW
  spectra (see, e.g., Figure~10 of Paper~II, for SN~1991bg).}.
Overall, the lines in Figure~4 provide a good overview of the
interpretation, at the SYNOW level, of the distribution of the SNe in
this plot.  However, the true nature of SN~Ia spectra in the region of
the 5750\ang\ and 6100\ang\ absorptions may be more complicated, in
some cases involving S~II, Fe~II, and Fe~III as well as Si~II (Bongard
et~al. 2008).  This may be why the positions of some SSs in Figure~4
are not accounted for by SYNOW spectra that contain only lines of
Si~II.\footnote{Lines for very low $v_e$ values (very steep
  optical--depth gradients), $\la500$ \kms, fall farther to the left
  in Figure~4, near most of the SSs, but the synthetic line profiles
  are not good matches to the observed ones.}.

On the basis of five spectroscopic and photometric parameters other
than $W(5750)$ and $W(6100)$, Benetti et~al. (2005) assigned 26 SNe~Ia
to three groups: faint; high temporal velocity gradient (HVG); and low
temporal velocity gradient (LVG).  As discussed in Paper~II, apart
from a few borderline exceptions their faint group corresponds to our
CLs; their HVG group corresponds to our BLs; and their LVG group
includes both our CNs and SSs.  Our understanding of the
correspondence between our groups and those of Benetti et~al., as
discussed in Paper~II, has not changed.

\section{CORE NORMALS}

The number of CNs has increased from 7 in Paper~II to 15 now.  The new
ones are SN 1997dt, SN~1998V, SN~1998eg, and SN~2000fa (Matheson
et~al. 2008), SN~2003cg (Elias--Rosa et~al. 2006), SN~2003du
(Stanishev et~al. 2007a), SN~2004S (Krisciunas et~al. 2007), and
SN~2005cf (Garavini et~al. 2007; Wang et~al. 2009).

By definition, at maximum light CNs have very similar spectra, except
for differing strengths of high velocity (HV) Ca~II features that form
above 20,000 \kms.  Among the new ones SN~2004S and SN~2005cf, like
SN~2001el of Paper~II, have strong HV Ca~II at maximum
light\footnote{A spectrum of Tycho Brahe's SN~1572 has been obtained
  by means of a light echo.  Krause et~al. (2008) showed that it is
  very similar to the time--integrated spectrum of the CN SN~2001el,
  and that of the CN SN~1994D except for the strong HV Ca~II
  absorption of SN~1572 and SN~2001el.  Thus Tycho's SN appears to
  have been a CN SN~Ia with strong HV Ca~II absorption.  Krause
  et~al. pointed out that by obtaining additional light--echo spectra
  it may be possible to construct a stereoscopic view of SN~1572,
  which would shed light on the geometry of the HV matter.}.  CNs are
somewhat more diverse at 1 week premaximum and much more diverse at
early times (Paper~III).

At postmaximum epochs CNs remain impressively homogeneous.  In
Paper~IV we attempted to illustrate the high degree of homogeneity of
CNs at 3 months postmaximum by plotting the logarithm of the flux with
vertically displaced spectra.  A better illustration of the homogenity
is presented in Figure~6, which overplots the (non--logarithmic) flux
spectra of the 6 CNs of the 3 month postmaximum sample.  Considering
observational error and that the epochs of the spectra range from
day~+83 to day~+91, the homogeneity from 4000\ang\ to 7000\ang\ is
remarkable and as argued in Paper~IV is incompatible with the usual
assumption that the spectra at these epochs (and even earlier) consist
of optically thin emission lines.  Instead, the blue part of the
spectrum is dominated by resonance scattering features of permitted
Fe~II lines and the Na~I feature.  To the red of 7000~\AA\ the
confusion is due to noise in the SN~1996X spectrum, to the differing
strengths and amounts of deredshifting of the 7600~\AA\ terrestrial
absorption, and to warping of the ends of some of the spectra by the
local normalization process.  Nevertheless, we cannot exclude that
some of the differences are real, perhaps due to differing strengths
of [Ca~II] emission lines in this region of low permitted line
opacity.  To the red of the Ca~II IR3 absorption near 8250\ang, the
spectrum may consist mainly of optically thin emission lines, as
usually assumed.  However, the overall shape of spectra modelled with
only optically thin forbidden lines will differ from the shape of
spectra modelled with forbidden and permitted lines, because the
optically thick permitted lines in the blue will transfer flux to the
red.

\section{BROAD LINES}

The number of BLs has increased from 7 in Paper~II to 20 now.  The new
ones are SN~1997do, SN~1998dh, SN~1998ec, SN~1999cc, SN~1999cl,
SN~1999ej, SN~1999gd, SN~1999gh, and SN~2000B (Matheson et~al. 2008),
SN~2002dj (Pignata et~al. 2008), SN~2004dt (Altavilla et~al. 2007),
and SN~2006X (Wang et~al. 2008).  Maximum light spectra are not
available for SN~1997do, SN~1999gh, and SN~2000B, but it is clear from
their 1 week postmaximum spectra that they should be classified as BL.

At maximum light BLs seem to have the same spectral features as CNs,
but the \W610 values are higher.  Some BLs are not very different from
CN, so there is no reason to suppose a lack of continuity from CNs to
these mild BLs.  Extreme BLs have very high \W610 values but not high
$W(5750)$ values, because (in SYNOW analyses) the Si~II optical depth
gradient is low ($v_e$ is high) and the optical depth at the
photosphere of \lam6355 is significant but not very high.  This causes
the \lam6355 absorption to extend over a wide velocity range, while
the optical depth of the weaker \lam5972 transition is too low to
produce significant absorption.

Although Figure~2 does not support any lack of continuity among BLs,
it is clear that there is no single parameter sequence from CNs to
extreme BLs.  For example, the comparison in Figure~7 of the maximum
light spectra of three extreme BLs (SN~2002bf, SN~2004dt, and
SN~2006X) that are clustered at the lower right of Figure~2, shows
that although the three have similar very broad 6100~\AA\ absorptions,
SN~2004dt strongly differs from SN~2002bf and SN~2006X in several
respects.  The most obvious one is that SN~2004dt lacks strong
absorption in the range 4800~\AA\ to 5100~\AA, which in SYNOW analyses
is attributed to HV Fe~II (Paper~II).  SN~2002bf and SN~2006X do have
strong HV Fe~II absorption, with the corresponding emission partially
filling in their S~II absorptions and making them appear weaker than
in SN~2004dt.

At 1 week premaximum, BLs are distinguishable from CN, but at the
earliest times the CN diversity is such that there is some overlap
between CN and BL.  At one week postmaximum, strong HV Fe~II and
partially filled in S~II remains a distinguishing characteristic of
most extreme BLs (see Figures~5 and 6 of Paper~IV), but mild BLs are
barely distinguishable from CN.  By three weeks postmaximum, even
extreme BLs are similar to CN.

\section{COOLS}

The number of CLs at maximum light has increased from 5 in Paper~II to
11 now.  The new ones are SN~1998bp, SN~1998de, SN~2000cn, and
SN~2000dk (Matheson et~al. 2008), SN~2004eo (Pastorello et~al. 2007),
and SN~2005bl (Taubenberger et~al. 2008).  A maximum light spectrum is
not available for SN~2000cn, but it is clear from its 1 week
premaximum and 3 weeks postmaximum spectra that it should be
classified as CL.

When Paper~II was written, the sample of maximum light CLs consisted
of SN~1991bg and two other extreme CLs, all with strong Ti~II
absorption; one less extreme CL, SN~1986G, with weaker Ti~II; and one
marginal CL, SN~1989B, with no identifiable Ti~II absorption.  Extreme
SN~1991bg--like CLs have at times been regarded as a distinct SN~Ia
subgroup, but in Paper~II we argued that there may be continuity from
CNs through SN~1986G to SN~1991bg.  It is interesting that among the
five new maximum light CL spectra SN~2005bl is SN~1991bg--like,
SN~1998de resembles SN~1986G, and three others (SN~1998bp, SN~2000dk,
and SN~2004eo) are very similar to each other and similar to SN~1986G
except for a lack of identifiable Ti~II absorption.  Figure~8 compares
SN~1998bp, SN~2000dk, and SN~2004eo to SN~1986G.  In LTE the Ti~II
strength increases rapidly with decreasing temperature (Hatano
et~al. 1999), so SN~1986G may just be slightly cooler than the others
of Figure~8.  Differences in titanium abundance may also be involved
(Taubenberger et~al. 2008).  The existence of several SNe (SN~1998de,
in addition to those of Figure~8) that resemble the intermediate CL
SN~1986G strengthens the case for continuity (although probably not a
single parameter sequence) from CN to extreme SN~1991bg--like CLs.

In Figure~2 SN~1989B is in the domain of BLs.  Nevertheless, in
Paper~II SN~1989B was included with CLs on the grounds that at maximum
light its features are better resolved than those of BLs and that its
6100\ang\ absorption is almost identical to that of the CL SN~1986G.
The mild CL SN~2004eo now provides support for associating SN~1989B
with CLs.  Pastorello et~al. (2007) refer to SN~2004eo as a
transitional object, between the LVG, HVG, and FAINT groups of Benetti
et~al. (2005). In our classification scheme, SN~2004eo is a mild CL
(Figure~8), and in many respects the spectra of SN~1989B resemble
those of SN~2004eo; for example, at 1 week premaximum the spectra are
very similar (Figure~9).  The reason for the weaker
5750\ang\ absorption of SN~1989B at maximum light is not known.

At 1 week premaximum CLs are distinguishable from CN.  There are no
CLs in the early sample.  At 1 week and even 3 weeks postmaximum, CLs
other than SN~1989B and SN~2004eo remain distinguishable from CN.

\section{SHALLOW SILICONS}

The number of SSs at maximum light has increased from 5 in Paper~II to
20 now.  The new ones are SN~1998ab, SN~1998es, SN~1999dq, SN~1999gp,
and SN~2001V (Matheson et~al. 2008), SN~1999ac (Garavini et~al. 2005),
SN~2000E (Valentini et~al. 2003), SN~2003fg (Howell et~al. 2006),
SN~2005cg (Quimby et~al. 2006), SN~2005hj (Quimby et~al. 2007),
SN~2005hk (Stanishev et~al. 2007b), and SN~2006gz (Hicken
et~al. 2007).  Maximum light spectra are not available for SN~1998ab
and SN~2001V, but it is clear from their 1 week premaximum spectra
that they should be classified as SS.

The SSs are a motley collection.  At maximum light, among the new
ones, SN~2005cg and SN~2005hj are not very far from CN; SN~1998ab is
not far from the extreme SS SN~1991T; SN~1999ac is mildly peculiar in
its own ways (Garavini et~al. 2005; Phillips et~al. 2006; Paper~IV);
SN~2003fg apparently has C~II absorptions and is suspected of
super--Chandrasekhar mass ejection (Howell et~al. 2006; Jeffery
et~al. 2008); SN~2006gz clearly has C~II absorptions and may also be
super--Chandrasekhar (Hicken et~al. 2007); SN~2005hk is a near clone
of SN~2002cx, both being quite distinct from typical SNe~Ia (Li
et~al. 2003; Jha et~al. 2006; Phillips et~al. 2007; Stanishev
et~al. 2007b; Sahu et~al. 2008).  In general, SSs are distinguishable
from CN at premaximum epochs and at 1 week postmaximum, but not at 3
weeks postmaximum.

Four new SSs not mentioned in the preceeding paragraph --- SN~1998es,
SN~1999dq, SN~1999gp, and SN~2001V are especially interesting not for
their individuality but for their similarity to each other, and to
SN~1999aa.  This is illustrated in Figure~10 for the maximum light
spectra of four of these (a maximum light spectrum is not available
for SN~2001V).\footnote {Rest et~al. (2008) obtained a light--echo
  spectrum of SNR 0509-67.5 in the Large Magellanic Cloud and showed
  that it is similar to the time--integrated spectra of SN~1998es,
  SN~1999aa, and SN~1999dq.  Thus the LMC remnant appears to be that
  of a 1999aa--like SS.}  Figure~11 is similar but for four SNe at 1
week premaximum. (SN~1999gp does not have a spectrum at this epoch but
SN~2001V does.)  Figure~12 shows the 3 week postmaximum spectra of all
five.  When the SN~1999aa-likes of the 3 week postmaximum sample are
compared to the well time--sampled series of SN~1994D spectra studied
in Paper~I, the SN~1999aa-likes are most like the day~+14 spectrum of
SN~1994D.  The ratio of days past maximum light, 19 for SN~1999aa to
14 for SN~1994D is 1.36, the same as the ratio of the B--band
light--curve stretch parameters of 1.143 for SN~1999aa and 0.838 for
SN~1994D (Takanashi et~al. 2008).  The possibility that CNs and SSs
age spectroscopically at the same rate that they decline
photometrically will be explored in a separate paper.

\section{QUANTITATIVE SPECTRAL SUBCLASSIFICATION OF TYPE Ia SUPERNOVAE}

A quantitative subclassification of SN~Ia spectra, preferably based on
a single spectrum obtained near maximum light, would be useful to
concisely convey a sense of what sort of SN~Ia one is referring to.
Before 1991, no spectroscopic subclassification scheme was used for
SNe~Ia, although it was known that some spectra differed from the norm
(e.g., Phillips et~al. 1987 on SN~1986G and Branch 1987 on SN~1984A).
After the appearance of the ``powerful'' SN~1991T (Ruiz--Lapuente et
al. 1992; Filippenko et~al. 1992a; Phillips et~al. 1992) and the
``weak'' SN~1991bg (Filippenko et~al. 1992b; Leibundgut et~al. 1993;
Turatto et~al. 1996), SNe~Ia usually were said to be either normal, or
peculiar SN~1991T--like, or peculiar SN~1991bg--like (e.g., Branch,
Fisher, \& Nugent 1993).  The term SN~1999aa--like was introduced by
Li et~al. (2001b) and recently we also have the enigmatic
SN~2002cx--likes (Li et~al. 2003; Jha et~al. 2006).  It has become
clear (e.g., from Figure~2) that neither these few terms by themselves
nor any single spectroscopic parameter can do justice to the
multi--dimensional SN~Ia maximum--light spectroscopic diversity.

We suggest that specifying the $W(5750)$ and $W(6100)$ parameters
would be a useful quantitative spectroscopic subclassification.  Only
one spectrum near maximum light\footnote {One week is a bit too far
  from maximum light because plots of $W(5750)$ and $W(6100)$ for the
  1 week premaximum and 1 week postmaximum samples, although
  qualitatively like Figure~2, show a higher degree of overlapping
  domains than at maximum light.} is required, and an accurate
redshift, sometimes not available for high redshift SNe~Ia, is not
needed.  (On the downside, some high redshift spectra do not reach
restframe 6100\ang, and our present equivalent widths depend on a
by--eye choice of where to begin and end the equivalent width
integrations.  A more precisely defined method such as that of Bongard
et~al. (2006) would be required for strictly reproducible results.)
Specifying only the group name, without the equivalent widths, is not
sufficient because, for example, it does not distinguish between mild
BLs (which in the traditional sense would be termed ``normal'') and
extreme BLs.  Specifying the equivalent widths, without the group
name, is insufficent because, for example, the CL SN~1989B falls
within the domain of the BLs.  Thus SN~2002bf could be denoted
SN~2002bf (BL/11/175).  This would concisely convey useful information
to those familiar with the $W(5750)$ -- $W(6100)$ plane much as a
spectral type of B2 IV concisely conveys useful information about
Acrux.

\section{ABSOLUTE MAGNITUDES}

This series of papers has been confined almost entirely to optical
spectra, but it is natural to wonder about the absolute magnitudes of
the four spectroscopic groups.  For our present sample the problem is
that most host galaxies are not in the Hubble flow, which makes it
difficult to compile a consistent set of distances.  To estimate
absolute magnitudes we have used data from Reindl et~al. (2005) and
Tammann, Sandage, \& Reindl (2008), including their estimates of host
galaxy extinction.  Distances in order of preference are based on (1)
a Hubble constant of 65 km~s$^{-1}$~Mpc$^{-1}$ provided that host
galaxy recession velocity\footnote{For SN~1999ej we have used 4690
\kms, the mean velocity of the cluster in which it occurred (Jha,
Riess, \& Kirshner,~R.~P. 2007).} exceeds 2000 km~s$^{-1}$, (2)
Cepheids and/or the tip of the red giant branch, and (3) membership in
the Virgo, Fornax, or Ursa major clusters.  The resulting Phillips
relation (Phillips et~al. 1999), for the SNe to which one of these
distances can be assigned\footnote {Owing to high host galaxy
extinction, CN SN~1997dt, BL SN~1999cl, and CLs SN~1986G and SN~1989B
are excluded.}, is shown in Figure~13.  It is no surprise that on
average SSs tend to be slowly declining (low \dm15) and bright, and
that CLs tend to be rapidly declining and faint.  Figure~13 also shows
that for this sample, BLs tend to have, on average, faster decline
rates than CNs and they may contribute substantial scatter to the
Phillips relation --- perhaps not surprising in view of the
considerable spectroscopic diversity among BLs.  Most importantly, the
spectroscopically homogeneous CNs may prove to be excellent standard
candles: for this small sample of 9 CNs, $\bar{M_B} = -19.48,
\sigma(M_B) = 0.14$, with no apparent dependence of $M_B$ on \dm15.
The dispersion is no larger than the characteristic uncertainty in
$M_B$.  This is tantalizing, but the absolute--magnitude homogeneity
of CNs must be tested with data on SNe in the Hubble flow, such as
that forthcoming from the Supernova Factory (Bailey et~al. 2008).

Figure~14 shows $M_B$ plotted against $W(5750)$.  The sample is the
same as in Figure~13 except that owing to the lack of a maximum light
spectrum SN~1997br is missing.  The 22 SNe having $W(5750) <
22$\ang\ have $M_B = -19.50, \sigma(M_B)=0.17.$ Evidently a maximum
light spectrum, before a measurement of \dm15\ is possible, could
eliminate much of the absolute magnitude scatter.

\section{SUMMARY}

By construction, core normals have a high degree of spectroscopic
similarity.  Extending away from CNs in Figure~2 are three ``arms'',
which we call broad line, cool, and shallow silicon.  These names are
descriptive of the extreme members of the groups, but it is important
to emphasize that the new data that have become available for this
paper reinforce the impression that SNe~Ia have a continuous
distribution of spectroscopic properties rather than consisting of
physically discrete subgroups.  (SN~2002cx--likes remain an apparent
exception.  Valenti et~al. 2009 suggest that they are core--collapse
events, and should be classified as SNe Ib/c.) Some BLs and SSs differ
only mildly from CN and now, with several new CLs resembling the
moderate CL SN~1986G, it seems likely that continuity also extends
from CNs to extreme CLs.  However, neither BLs nor SSs consist of
single--parameter sequences extending from mild to extreme members,
and the same appears to be true of CLs. It is not yet clear whether,
apart from CNs, there is additional ``clumping'' in the spectroscopic
properties; e.g., it will be interesting to see whether the apparent
relatively high frequency of SN~1999aa--likes proves to be real or
coincidental.

The quantitative spectroscopic subclassification suggested in \S 8
would be an improvement over anything in current use, but even that
scheme cannot be expected to encompass the complex multi--dimensional
spectroscopic diversity among SNe~Ia.  More observational data are
needed to further illuminate it, and more work on explosion models is
needed to understand it.

We are grateful to Rollin Thomas and Xiaofeng Wang for helpful
comments and, as always, to all observers who provided spectra.  This
work has been supported by NSF grants AST 0506028 and AST 0707704,
NASA LTSA grant NNG04GD36G, and DOE grant DEFG02-07ER41517.

\clearpage     

\begin{figure}
\includegraphics[width=.8\textwidth,angle=0]{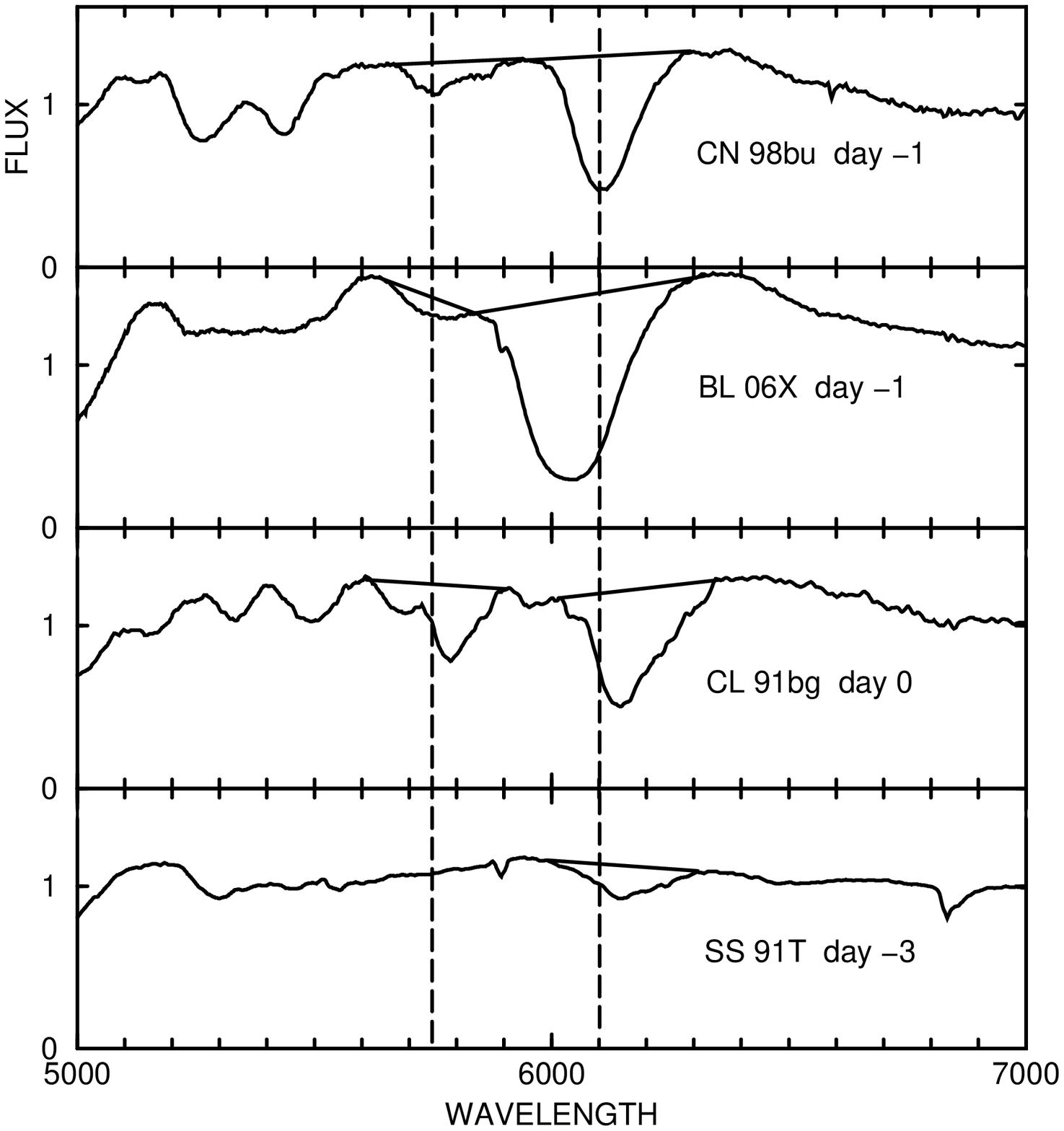}
\caption{Comparison of maximum light spectra of the CN SN~1998bu, the
  extreme BL SN~2006X, the extreme CL SN~1991bg, and the extreme SS
  SN~1991T.  Vertical dashed lines are at 5750\ang\ and 6100\ang.
  Straight solid lines show how $W(5750)$ and $W(6100)$ were measured.
 (For clarity, the line for $W(5750)$ in SN~1991T is not shown.)}
\end{figure}

\begin{figure}
\includegraphics[width=.8\textwidth,angle=270]{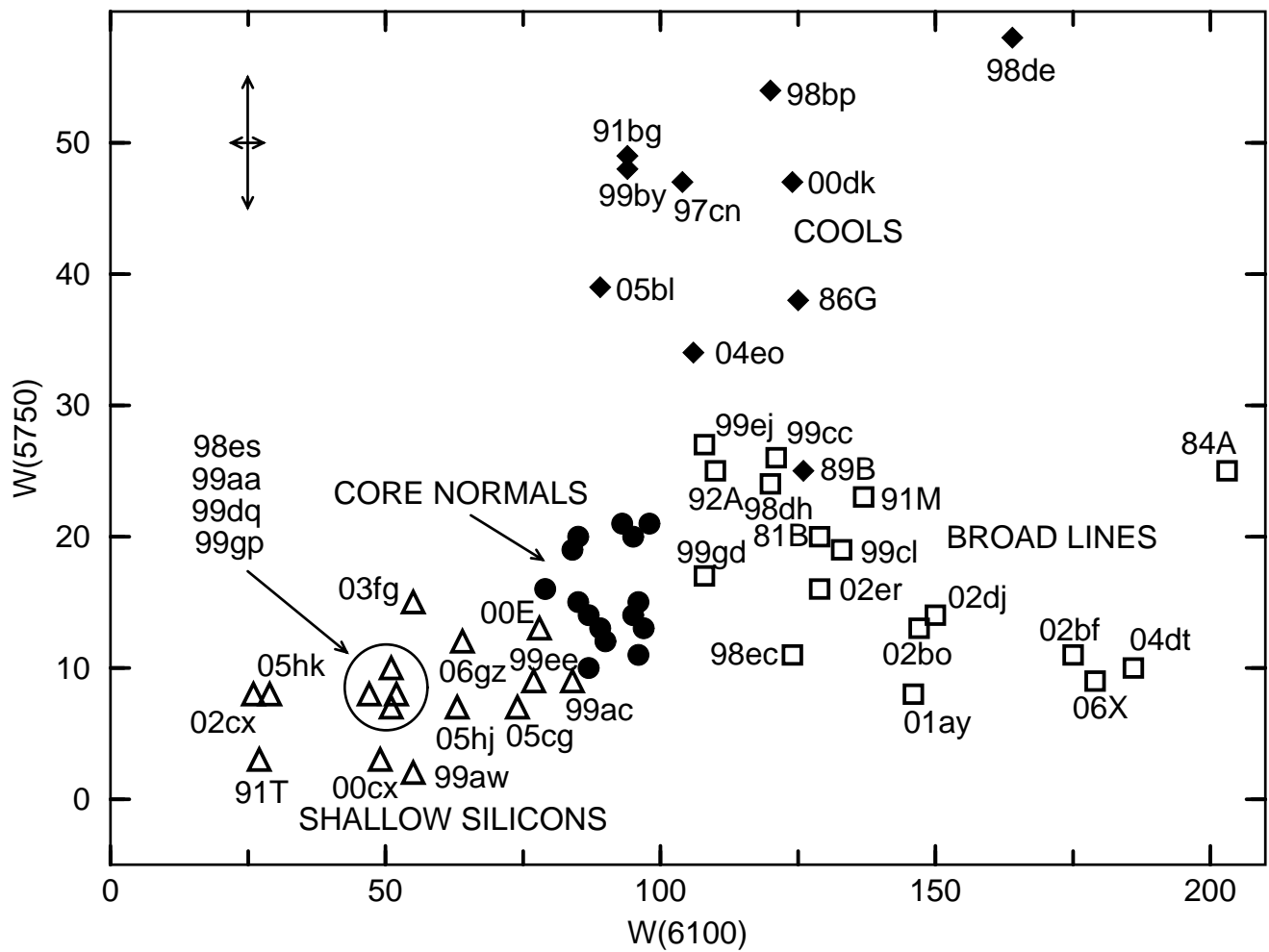}
\caption{$W(5750)$ plotted against $W(6100)$.  CNs are shown as filled
  circles, BLs as open squares, CLs as filled diamonds, and SSs as
  open triangles.  Characteristic uncertainties of 3~\AA\ in $W(5750)$
  and 5~\AA\ in $W(6100)$ are shown.  For clarity individual CNs are
  not labeled.}
\end{figure}

\begin{figure}
\includegraphics[width=.8\textwidth,angle=270]{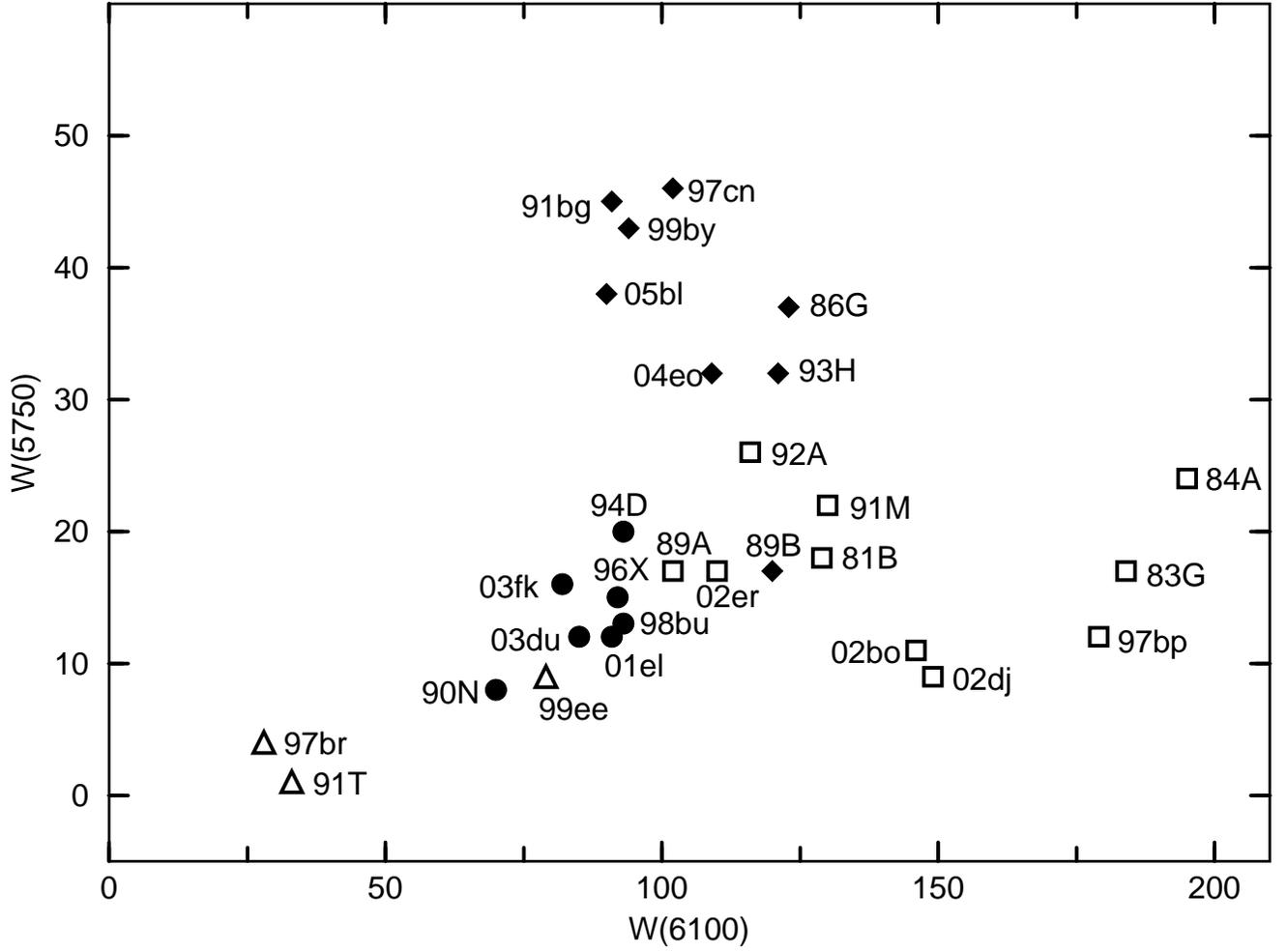}
\caption{Like Figure~2 but for equivalent widths measured by Hachinger
  et~al. 2006.}
\end{figure}

\begin{figure}
\includegraphics[width=.8\textwidth,angle=270]{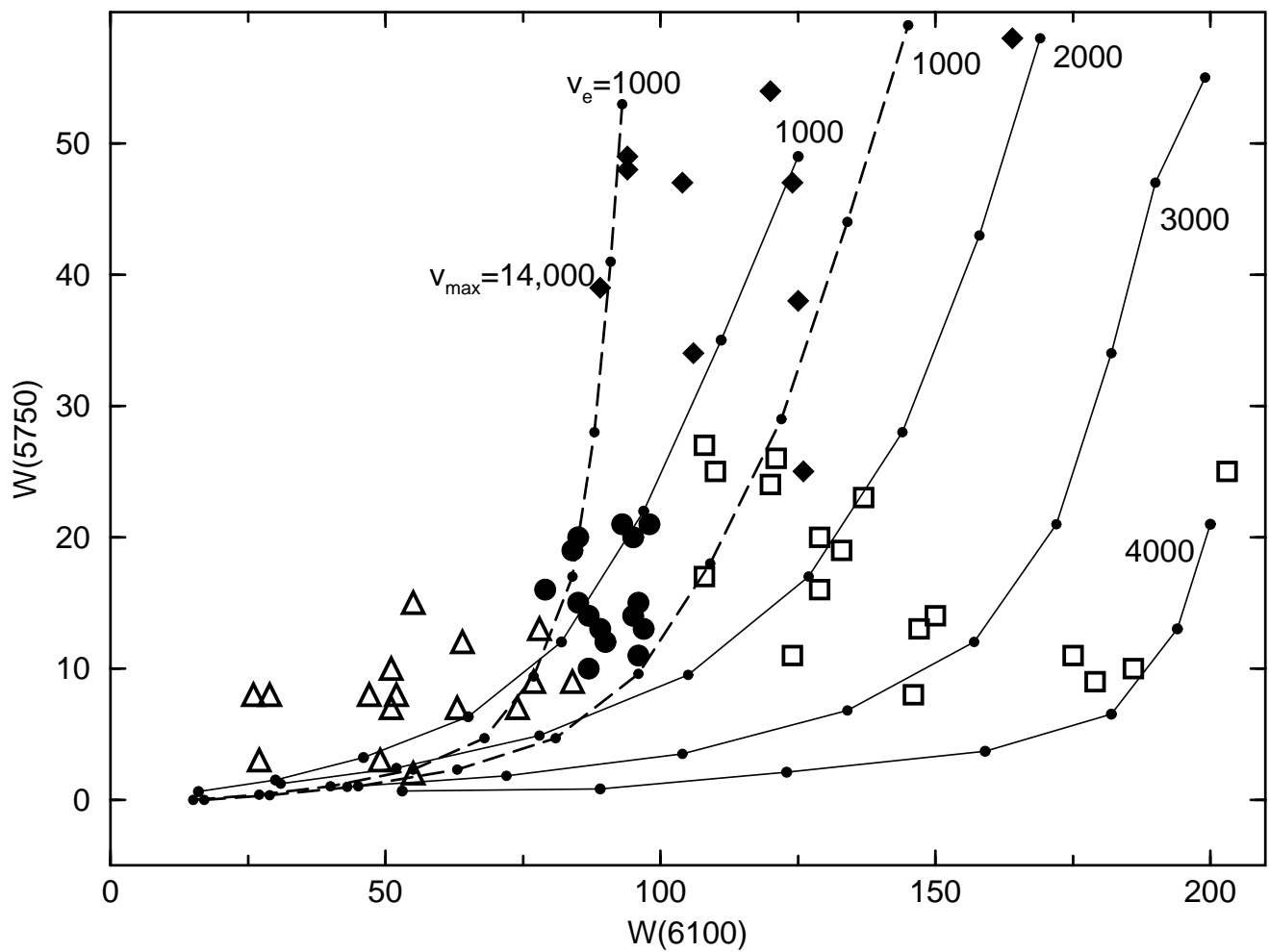}
\caption{Like Figure~2 but with lines representing measurements of
  $W(5750)$ and $W(6100)$ from SYNOW synthetic spectra.  Solid lines
  are for $v_{phot}=12,000$ \kms, $T_{exc}=10,000$~K, and $v_e$ from
  1000 to 4000 \kms\ as indicated.  Dashed lines (relevant only for
  CLs) are for $v_{phot}=11,000$ \kms, $T_{exc}=7000$~K, and $v_e =
  1000$ \kms.  One of the dashed lines is for a maximum velocity
  $v_{max}=14,000$ \kms, the other is for no maximum velocity.  For
  each line, the optical depth at the photosphere of Si~II \lam6355
  begins at 0.5 (lower left point) and then increases by factors of
  2.}
\end{figure}

\begin{figure}
\includegraphics[width=.8\textwidth,angle=270]{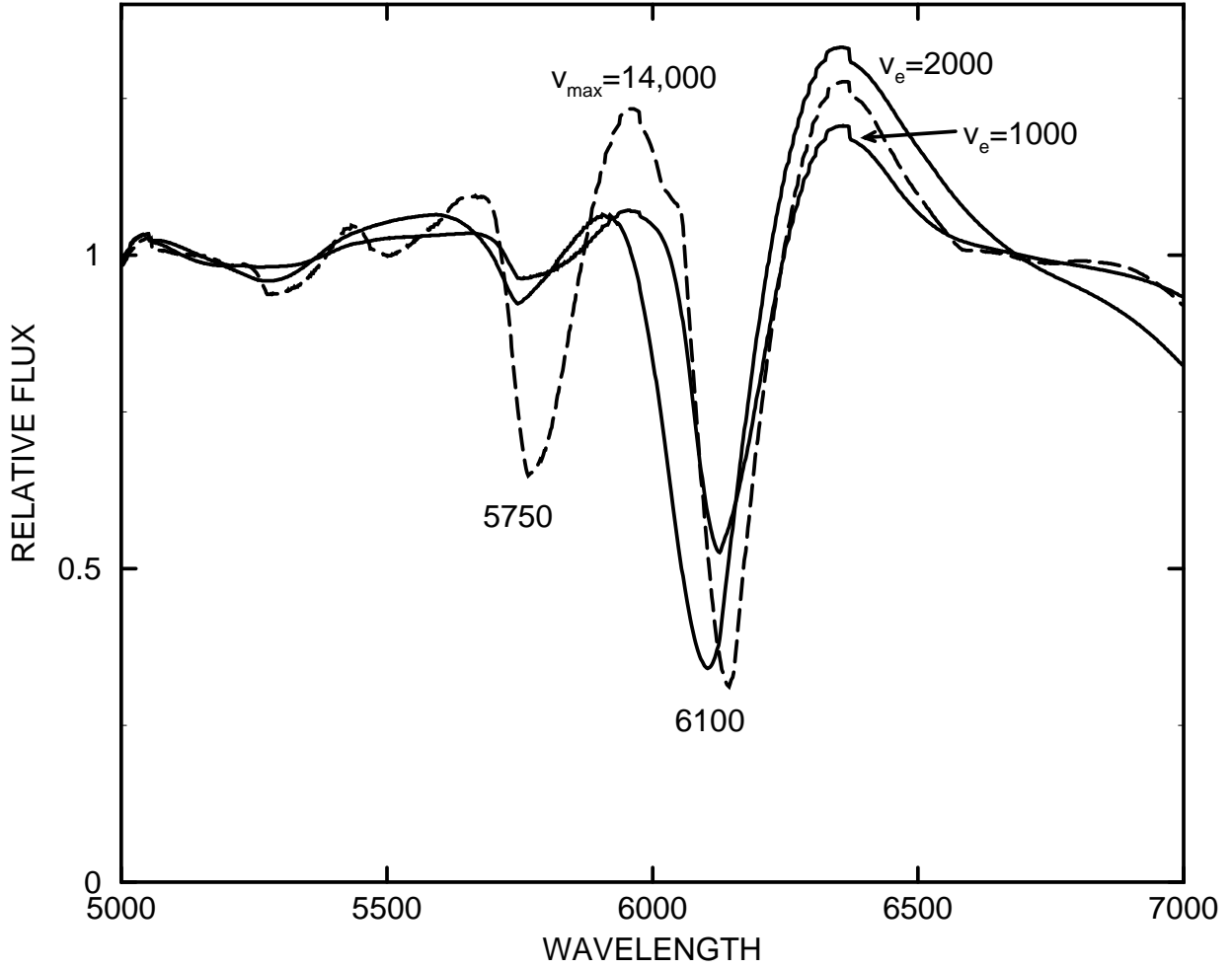}
\caption{Example SYNOW synthetic spectra for Si II.  Solid lines are
  for $v_{phot}=12,000$ \kms, $\tau=8$, $T_{exc}=10,000$~K, and
  $v_e=1000$ and 2000 \kms.  Dashed line (relevant only to CLs) is for
  $v_{phot}=11,000$ \kms, $\tau=100$, $T_{exc}=7000$~K, $v_e=1000$
  \kms, and $v_{max}=14,000$ \kms.  The maximum velocity together with
  a high optical depth allows the 5750\ang\ absorption to be strong
  without having the 6100\ang\ absorption excessively strong.}
\end{figure}

\begin{figure}
\includegraphics[width=.8\textwidth,angle=270]{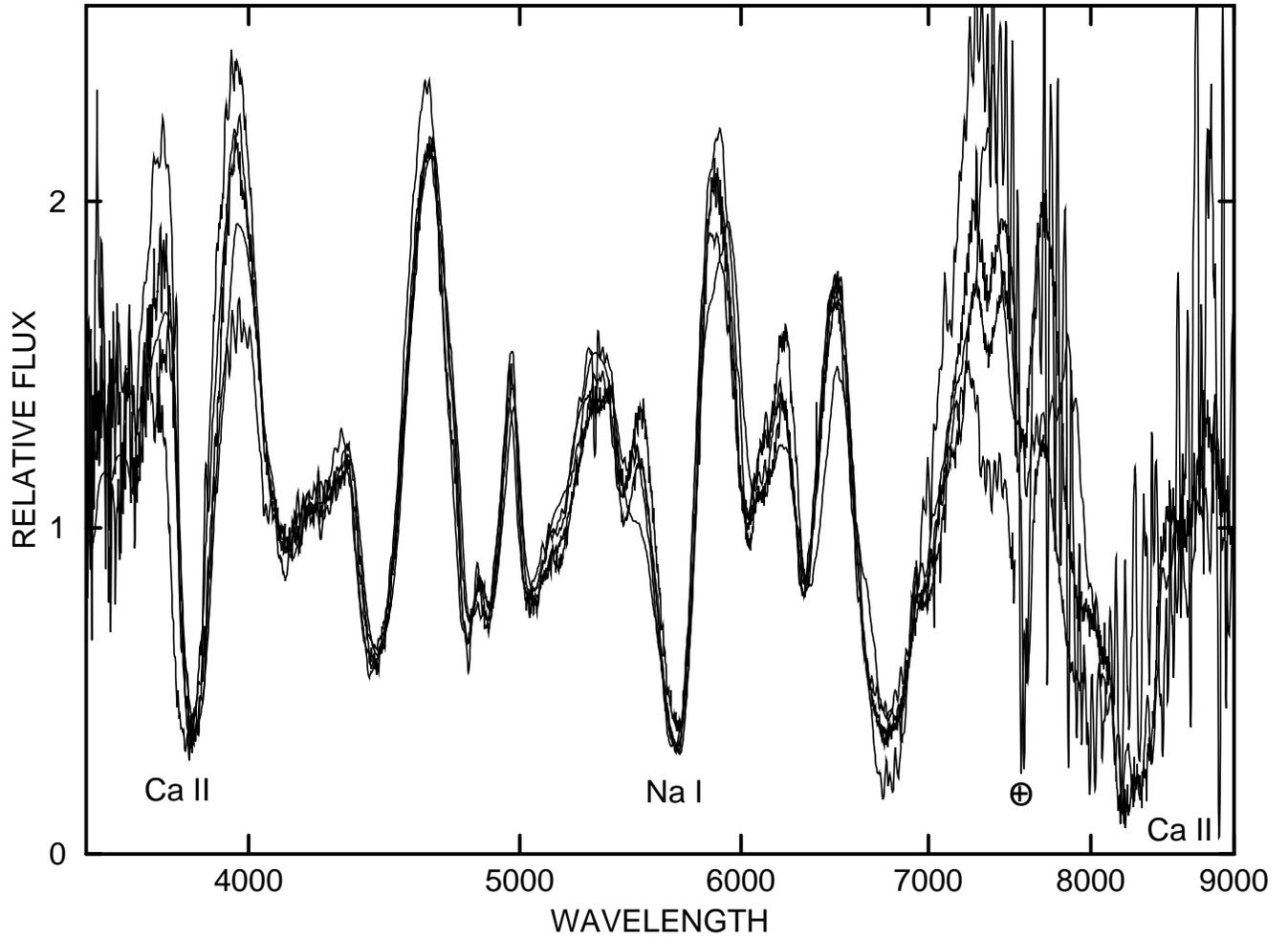}
\caption{Spectra of the six CNs (SN~1994D, SN~1994ae, SN~1996X,
  SN~1998aq, SN~2003du, SN~2005cf) of the 3 months postmaximum sample.
  Unlabeled absorption features between 4000\ang\ and 6500\ang\ are
  attributed to permitted lines of Fe~II (Paper~IV).}
\end{figure}

\begin{figure}
\includegraphics[width=.8\textwidth,angle=270]{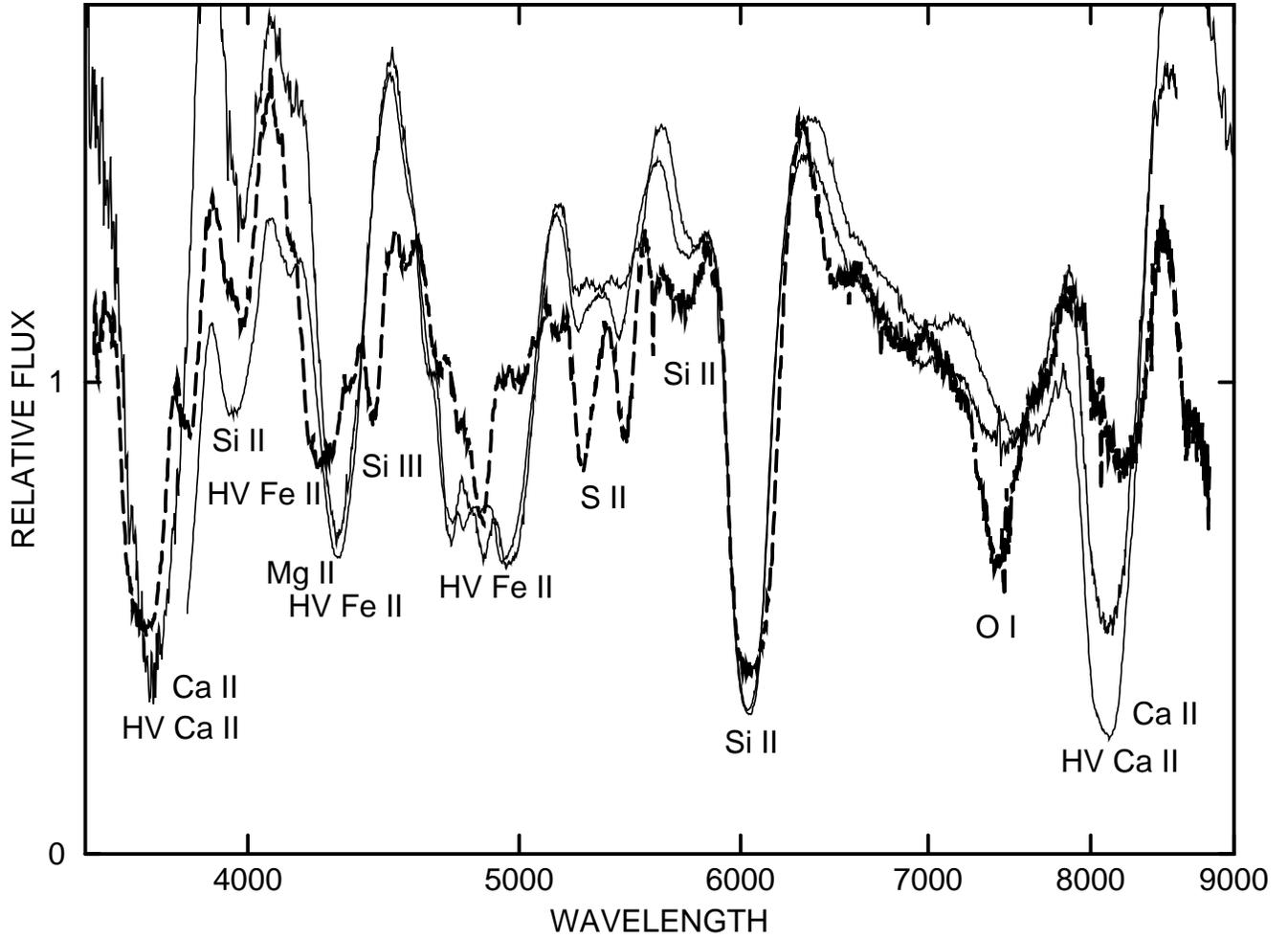}
\caption{Maximum light spectra of 3 extreme BLs: SN~2004dt (dashed
line), and SN~2002bf and SN~2006X (solid lines).  The 6100\ang\
absorptions are similar but SN~2004dt strongly differs from the other
two in several respects.}
\end{figure}

\begin{figure}
\includegraphics[width=.8\textwidth,angle=270]{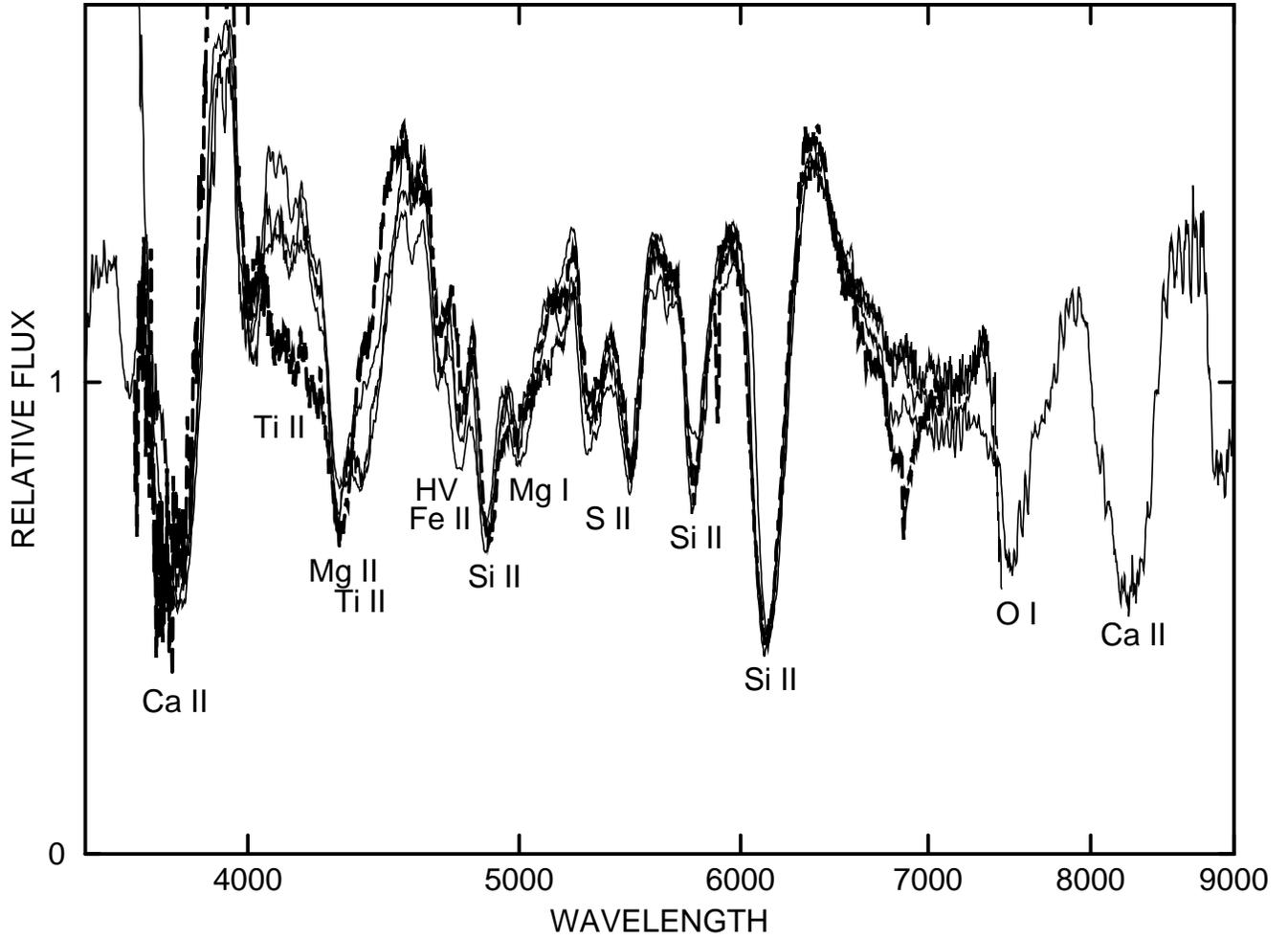}
\caption{Maximum light spectra of 3 CLs (SN~1998bp, SN~2000dk, and
SN~2004eo (solid lines), compared to SN~1986G (dashed line).  Line
identifications are for SN~1986G, which differs from the others in
having stronger absorption near 4200\ang, attributed mainly to Ti~II.}
\end{figure}

\begin{figure}
\includegraphics[width=.8\textwidth,angle=270]{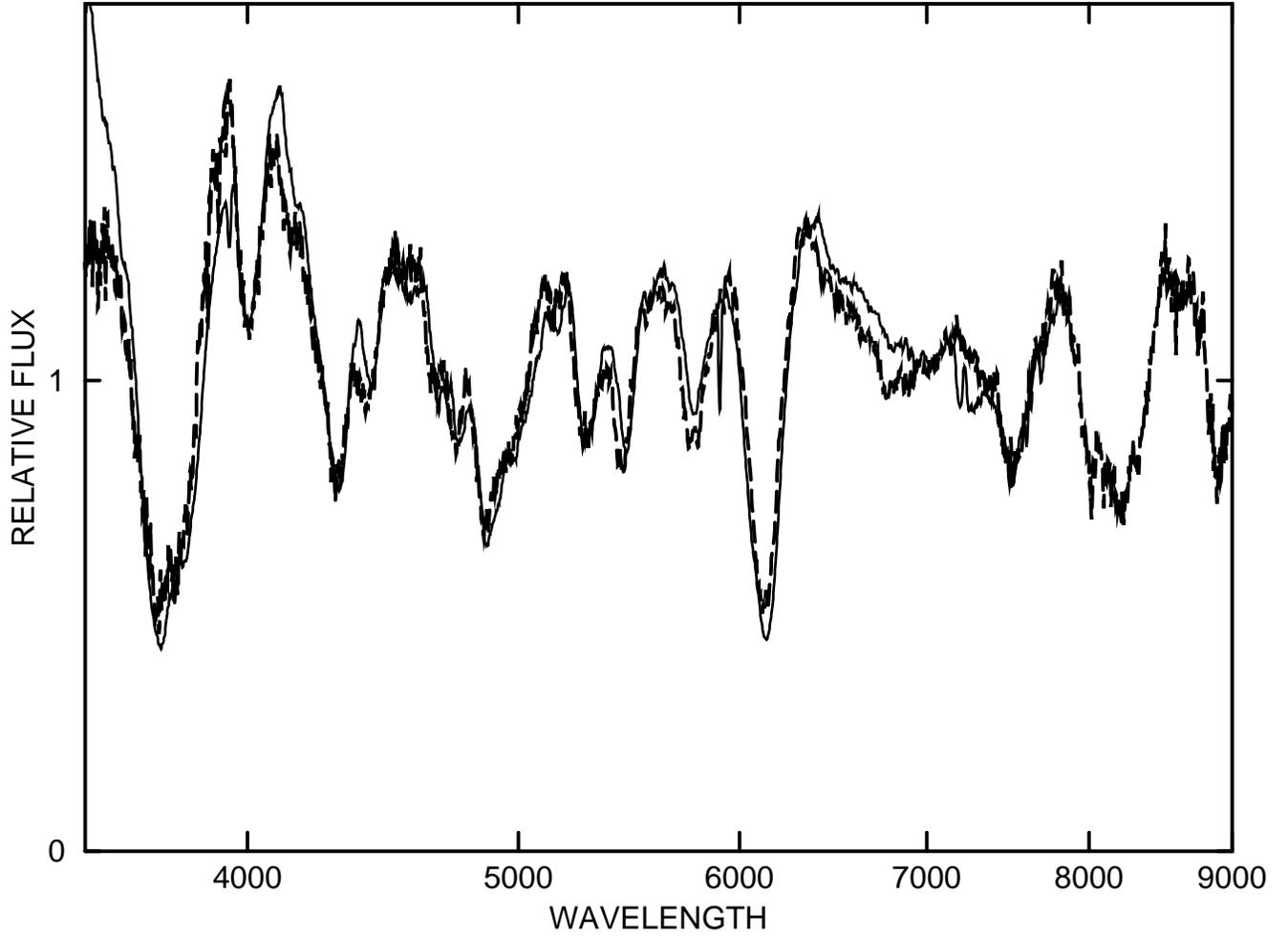}
\caption{One week premaximum spectra of the mild CLs SN~1989B (solid
  line) and SN~2004eo (dashed line).  The spectra are similar .}
\end{figure}

\begin{figure}
\includegraphics[width=.8\textwidth,angle=270]{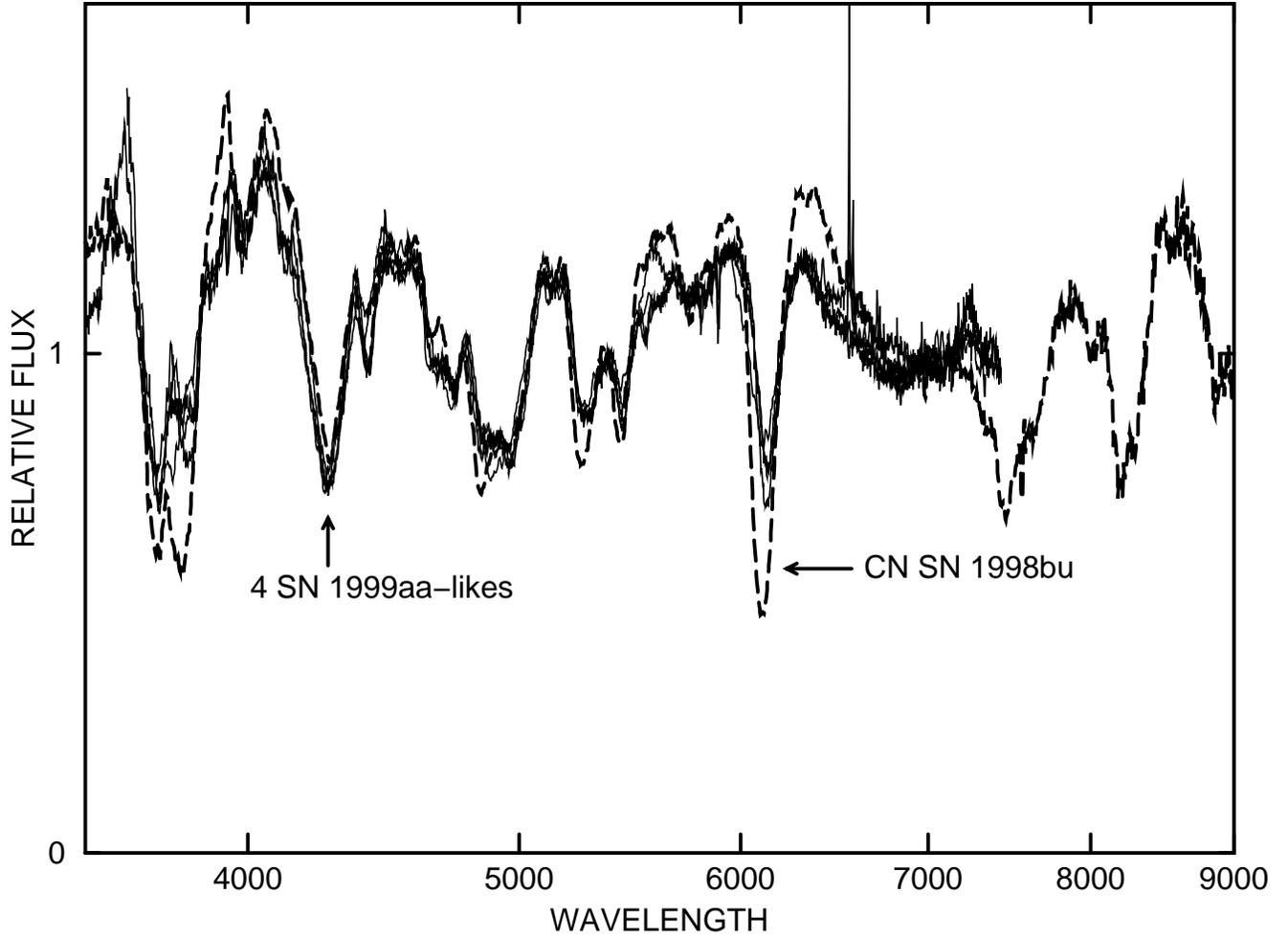}
\caption{Maximum light spectra of 4 SSs --- SN~1998es, SN~1999aa,
  SN~1999dq, and SN~1999gp (solid lines), compared to the CN SN~1998bu
  (dashed line).  The 4 SN~1999aa--likes are very similar to each
  other and distinguishable from SN~1998bu.}
\end{figure}

\begin{figure}
\includegraphics[width=.8\textwidth,angle=270]{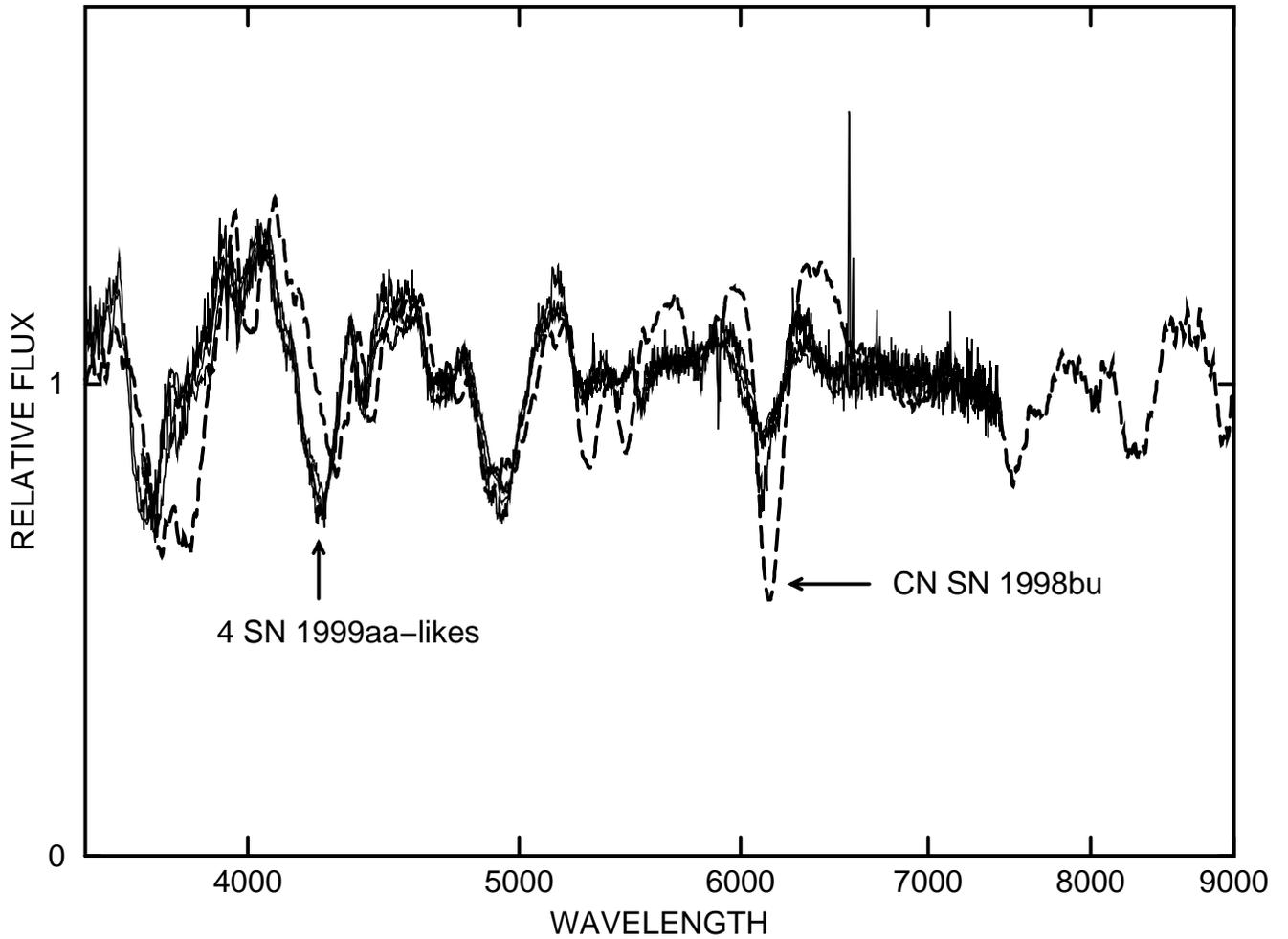}
\caption{One week premaximum spectra of 4 SSs --- SN~1998es,
  SN~1999aa, SN~1999dq, and SN~2001V (solid lines), compared to the CN
  SN~1998bu (dashed line).  The 4 SN~1999aa--likes are very similar to
  each other (although SN~2001V has a mildly deeper
  6100\ang\ absorption than the others) and distinguishable from
  SN~1998bu.}
\end{figure}

\begin{figure}
\includegraphics[width=.8\textwidth,angle=270]{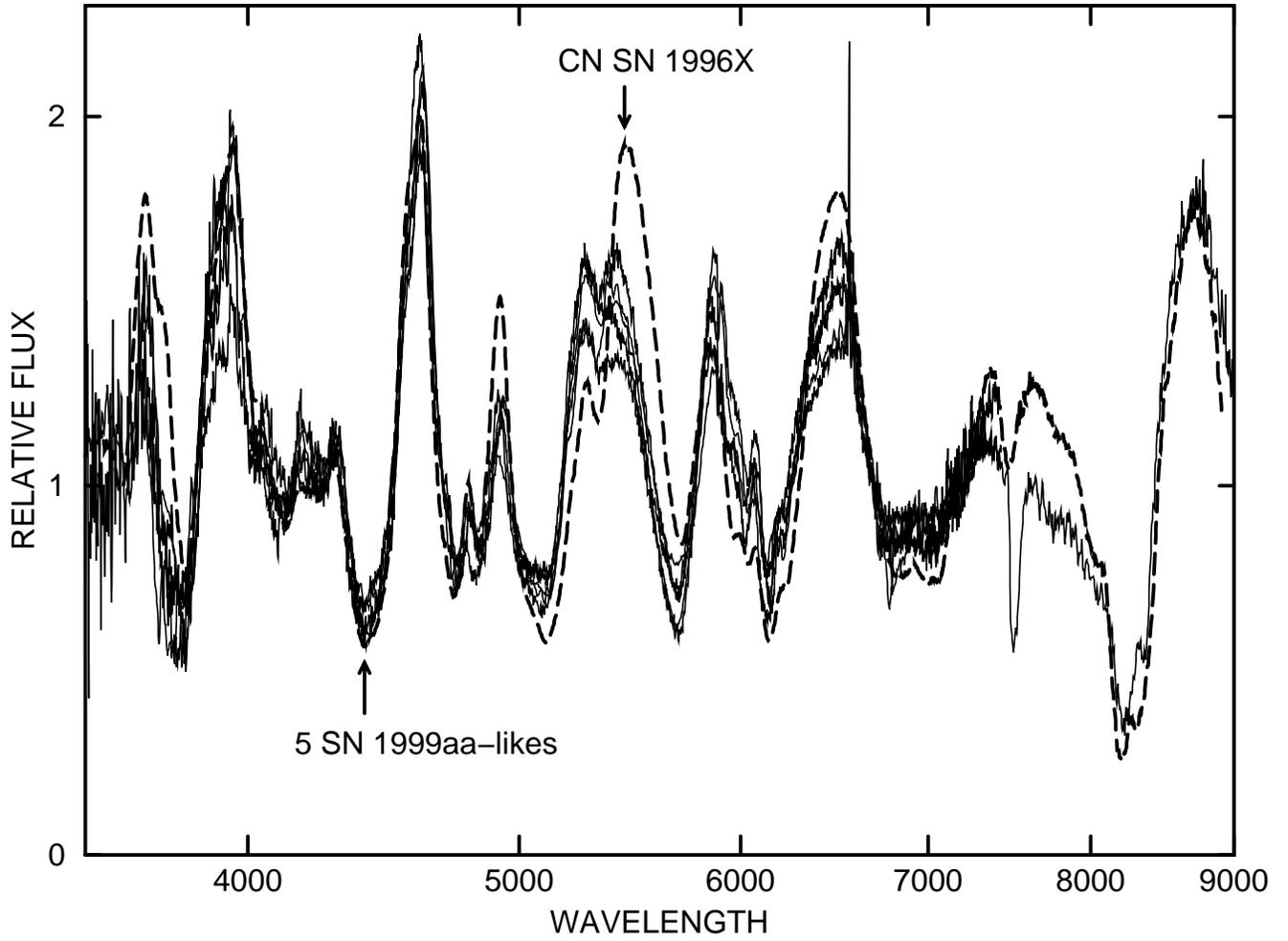}
\caption{Three week postmaximum spectra of 5 SSs --- SN~1998es,
  SN~1999aa, SN~1999dq, SN~1999gp, and SN~2001V (solid lines),
  compared to the CN SN~1996X (dashed line).  The 5 SN~1999aa--likes
  are very similar to each other and distinguishable from SN~1996X.}
\end{figure}

\begin{figure}
\includegraphics[width=.8\textwidth,angle=270]{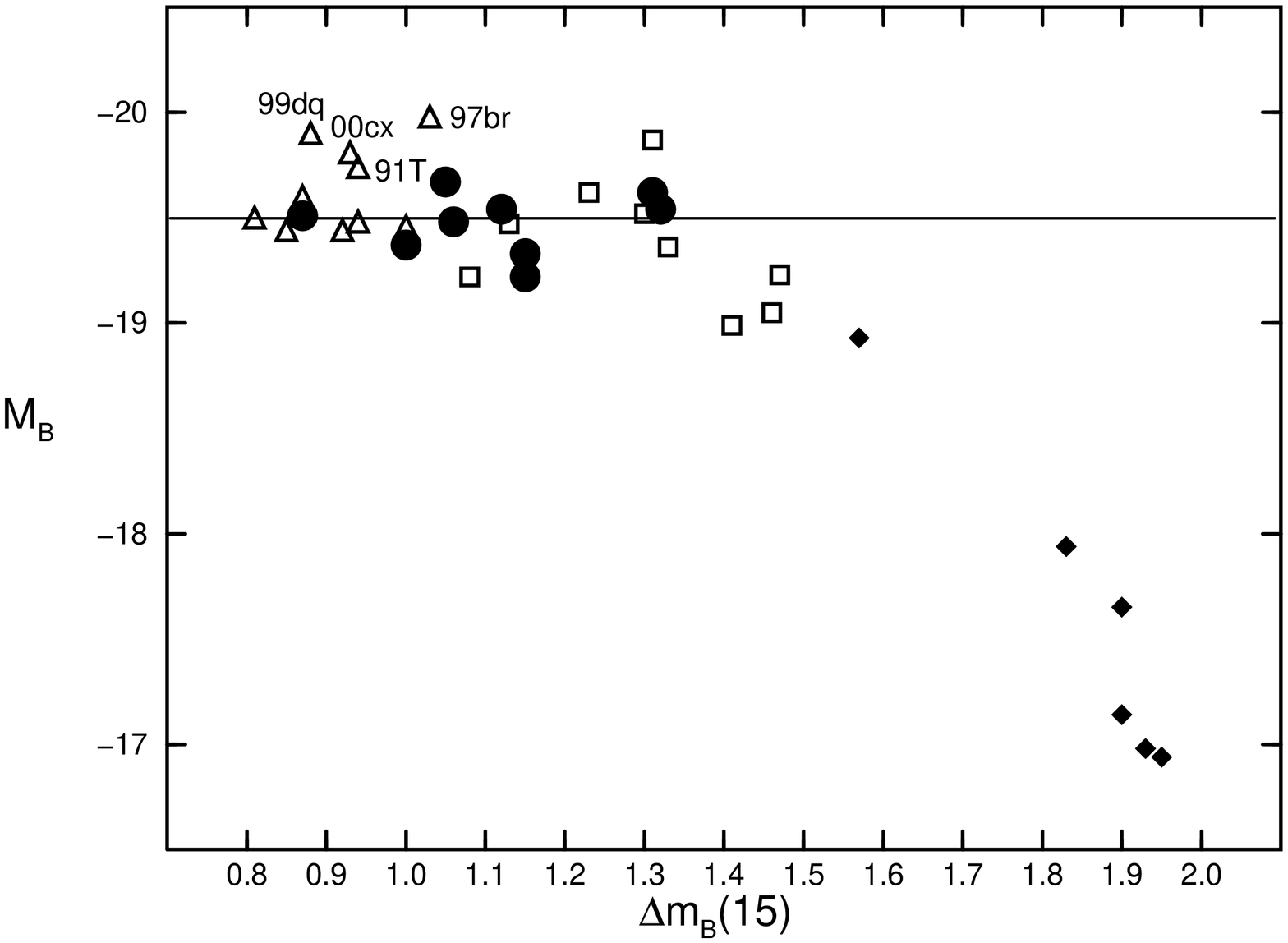}
\caption{The Phillips relation.  CNs are shown as filled circles, BLs
  as open squares, CLs as filled diamonds, and SSs as open triangles.
  The horizontal line is at $M_B = -19.5.$}
\end{figure}

\clearpage

\begin{figure}
\includegraphics[width=.8\textwidth,angle=270]{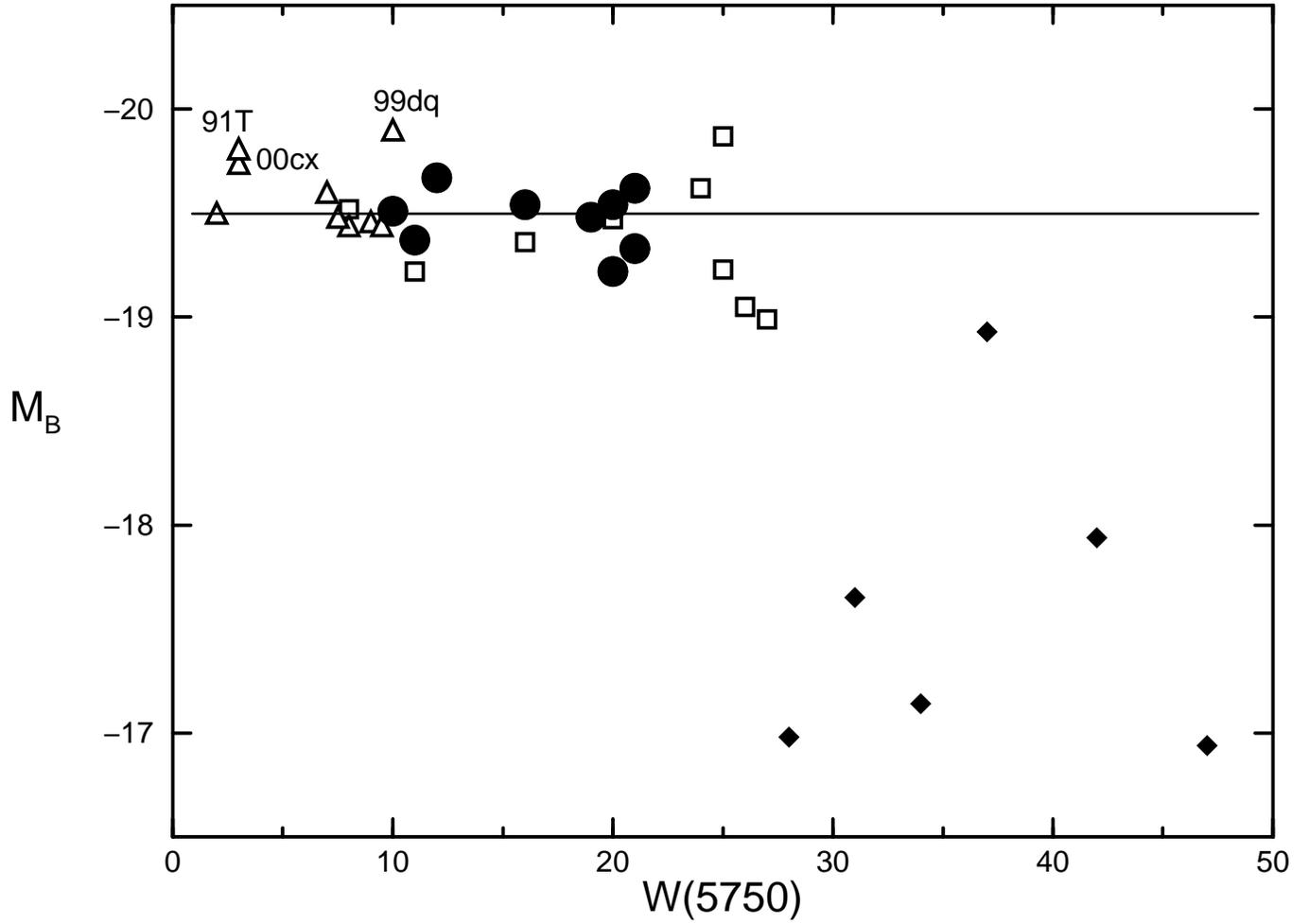}
\caption{$M_B$ plotted against $W(5750)$.  Symbols are as in
  Figure~13.  The horizontal line is at $M_B = -19.5.$}
\end{figure}
\clearpage

\begin{deluxetable}{lccccl}
%\tabletypesize{\scriptsize}
\rotate
\tablenum{1}
%\tablewidth{1000pt}
\setlength{\tabcolsep}{4pt}

\tablecaption{The SN Ia Sample}

\tablehead{ \colhead{SN} & \colhead{Epochs (days)} & \colhead{Galaxy}
& \colhead{$W(6100)$ (\AA)} & \colhead{$W(5750)$ (\AA)} &
\colhead{References} }

%\colhead{} & \colhead{(days)} & \colhead{} &
%\colhead{(\ang)} & \colhead{(\ang)} & \colhead{} }

\startdata

1981B BL & $-$2, 22, 93 & NGC 4536 & 129 & 20 & Branch et al. 1983\\

1984A BL & $-$7, $-$3, 8, 19 & NGC 4419 & 203 & 25 & Barbon et al. 1989\\

1986G CL & $-6$, $-$1, 21, 90 & NGC 5128 & 125 & 38 & $-6$: Phillips
et~al. 1987\\ & & & & & $-1$, 21, 90: Cristiani et al. 1992\\

1989B CL & $-$7, 0, 8, 19, 92 & NGC 3627 & 126 & 25 & Wells et al. 1994\\

1990N CN & $-14$, $-8$, $-$2, 7, 21 & NGC 4639 & 90 & 12 & $-14$,
$-8$, 7: Leibundgut et~al. 1991\\ & & & & & $-2$: Asiago SN Group\\ &
& & & & 21: Filippenko et~al. 1992b\\

1991M BL & 3, 81 & IC 1151 & 137 & 23 & 3: Gomez et~al. 1996\\& & & &
& 81: Gomez \& Lopez 1998\\

1991T SS & $-11$, $-7$, $-$3, 83 & NGC 4527 & 27 & 3 & $-11$, $-7$,
$-$3: Phillips et~al. 1992\\ & & & & & 83: A.~V.~Filippenko,
unpublished\\

1991bg CL & 0, 19, 91 & NGC 4374 & 94 & 49 & Filippenko et~al. 1992a\\

1992A BL & $-6$, $-$1, 6 & NGC 1380 & 110 & 25 & $-6$, $-$1:
P.~Challis, unpublished\\  & & & & & 6: Kirshner et~al. 1993\\

1994D CN & $-12$, $-8$, $-$1, 7, 19, 87 & NGC 4526 & 98 & 21 & $-12$,
$-8$, $-$1: Meikle et~al. 1996\\ & & & & & 7: Patat et~al. 1996\\ & &
& & & 87: Filippenko 1997\\

1994ae CN & 0, 89 & NGC 3370 & 87 &10 & 0: Howell \& Nugent 2004\\ & &
& & & 89 Bowers et~al. 1997\\

1996X CN & $-$2, 7, 22, 87 & NGC 5061 & 85 & 20 & Salvo et al. 2001\\

1997br SS & $-7$, 8 & ESO 576-G40 & ... & ... & Li et~al. 1999\\

1997cn CL & 3 & NGC 5490 & 104 & 47 & Turatto et~al. 1998\\

1997do BL & 8, 21 & UGC 3845 & ... & ... & Matheson et~al. 2008\\

1997dt CN & $-7$, 1 & NGC 7448 & 87 & 14 & Matheson et~al. 2008\\

1998V CN & 0 & NGC 6627 & 84 & 19 & Matheson et~al. 2008\\

1998ab SS & $-7$, 21 & NGC 4704 & ... & ... & Matheson et~al. 2008\\

1998aq CN & $-8$, 0, 7, 21, 91 & NGC 3982 & 79 & 16 & Branch et
al. 2003\\

1998bp CL & 0 & NGC 9495 & 120 & 54 & Matheson et~al. 2008\\

1998bu CN & $-6$, $-$1, 8 & NGC 3368 & 93 & 21 & $-6$, $-$1: Hernandez
et~al. 2000\\ & & & & & 8: Jha et~al. 1999\\

1998de CL & $-6$, 0 & NGC 252 & 164 & 58 & Matheson et~al. 2008\\

1998dh BL & $-7$, 0 & NGC 7541 & 120 & 24 & Matheson et~al. (2008\\

1998ec BL & $-1$, 21 & NGC 5948 & 124 & 11 & Matheson et~al. 2008)\\

1998eg CN & 0, 6, 20 & NGC 7391 & 95 & 20 & Matheson et~al. 2008\\

1998es SS & $-7$, $-1$, 20, 81 & NGC 3157 & 51 & 7 & Matheson
et~al. 2008\\

1999aa SS & $-11$, $-7$, $-1$, 6, 19, 81 & NGC 4469 & 63 & 14 & $-11$,
6, 19, 81: Garavini et~al. 2004\\ & & & & & $-7$, $-1$: Matheson
et~al. 2008\\

1999ac SS & $-15$, $-9$, $-2$, 8, 86 & NGC 6063 & 84 & 9 & $-15$,
$-9$, $-2$, 8: Garavini et~al. 2005\\& & & & & 86: Matheson
et~al. 2008\\

1999aw SS & 3 & $...$ & 55 & 2 & Howell \& Nugent 2004\\

1999by CL & $-$3, 7 & NGC 2841 & 94 & 48 & Garnavich et al. 2004\\

1999cc BL & 0, 23 & NGC 6038 & 121 & 26 & Matheson et~al. 2008\\

1999cl BL & $-7$, 0, 8 & NGC 4501 & 133 & 19 & Matheson et~al. 2008\\

1999dq SS & $-7$, 1, 6, 19 & NGC 976 & 51 & 10 & Matheson et~al. 2008\\

1999ee SS & $-7$, $-$2, 22 & IC 5179 & 77 & 9 & Hamuy et al. 2002\\

1999ej BL & 0 & NGC 495 & 108 & 27 & Matheson et~al. 2008\\

1999gd BL & 2 & NGC 5472 & 108 & 17 & Matheson et~al. 2008\\

1999gh BL & 7 & NGC 2986 & ... & ... & Matheson et~al. 2008\\

1999gp SS & $-1$, 7, 21 & ... & 52 & 8 & Matheson et~al. 2008\\

2000B BL & 8, 21 & NGC 2320 & ... & ... & Matheson et~al. 2008\\

2000E SS & $-6$, $-2$, 8 & NGC 6951 & 78 & 13 & Valentini et~al. 2003\\

2000cn CL & $-7$, 21 & NGC 7043 & ... & ... & Matheson et~al. 2008\\

2000cx SS & 2, 7, 20, 89 & NGC 524 & 49 & 3 & Li et al. 2001a\\

2000dk CL & 1 & NGC 5217 & 124 & 47 & Matheson et~al. 2008\\

2000fa CN & 1, 21 & NGC 6379 & 96 & 11 & Matheson et~al. 2008\\

2001V SS & $-13$, $-7$, 20 & NGC 4589 & ... & ... & Matheson et~al. 2008\\

2001ay BL & 0 & IC 4423 & 146 & 8 & Howell \& Nugent 2004\\

2001el CN & 1, 20 & NGC 1448 & 95 & 14 & Wang et al. 2003\\

2002bf BL & 3, 7 & $...$ & 175 & 11 & Leonard et al. 2005\\

2002bo BL & $-14$, $-8$, $-$1, 82 & NGC 3190 & 147 & 13 & Benetti et
al. 2004\\

2002cx SS & $-$1, 21 & ...  & 26 & 8 & Li et al. 2003\\

2002dj BL & $-$11, $-6$, $-3$, 23 & NGC 5018 & 150 & 14 & Pignata et
al. 2008\\

2002er BL & $-11$, $-$7, 0, 6, 20 & UGC 10743 & 129 & 16 & Kotak et
al. 2006\\

2003cg CN & $-$7, $-1$, 7, 23 & NGC 3169 & 96 & 15 & Elias--Rosa et
al. 2006\\

2003du CN & $-11$, $-$7, 0, 7, 84 & UGC 9391 & 85 & 15 & Stanishev et
al. 2007a\\

2003fg SS & 2 & ... & 55 & 15 & Howell et~al. 2006\\

2004S CN & 1, 8, 19 & ... & 89 & 13 & Krisciunis et al. 2007\\

2004dt BL & $-$8, $-1$, 20 & NGC 799 & 186 & 10 & Altavilla et
al. 2007\\

2004eo CL & 2, 7, 21 & NGC 6928 & 106 & 34 & Pastorello et
al. 2007\\

2005bl CL & $-$6, $-3$, 20 & NGC 4070 & 89 & 39 & Taubenberger et
al. 2008\\

2005cf CN & $-12$, $-$6, 0, 7 & ... & 97 & 13 & Garavini et
al. 2007\\ & & & & & 83: Wang et~al. 2009\\

2005cg SS & 0, 7 & ... & 74 & 7 & Quimby et
al. 2006\\

2005hj SS & $-$6, 0 & ... & 63 & 7 & Quimby et al. 2007\\

2005hk SS & 0, 21 & UGC 272 & 29 & 8 & Stanishev et al. 2007b\\

2006X BL & 1, 6, 98 & NGC 4321 & 179 & 9 & Wang et
al. 2007\\

2006gz SS & $-$14, $-2$, 7 & IC 1277 & 64 & 12 & Hicken et al. 2007\\

\enddata

\end{deluxetable}

\end{document}